# The 2008 Australian study of remnant data contained on 2$^{nd}$ hand hard disks: the saga continues

Craig Valli
SECAU – Security Research Centre
Edith Cowan University
c.valli@ecu.edu.au

Andrew Woodward
SECAU – Security Research Centre
Edith Cowan University
a.woodward@ecu.edu.au

**Abstract**

*This study looked for remnant data on enterprise level hard drives that were purchased through auctions. The drives were analysed for information, be it topical or formatted. In the event that drives were formatted, forensic tools were used to recover this data. This years study revealed a high level of not simply un-erased drives, but drives which contained information that related to critical infrastructure providers. That such a small sample size yielded such a high rate of un-erased drives is of considerable concern, and it may be necessary for the government to become involved.*

## INTRODUCTION

The issue of remnant data on electronic storage devices, and hard drives in particular, is one that has been covered by a variety of researchers over a number of years. The authors first started examining hard drives for remnant data on second hand drives in 2004 (Valli 2004), with the study continued in subsequent years (Valli & Jones, 2005; Jones *et al*, 2006; Valli & Woodward, 2007). Whilst other storage media has the potential to contain remnant data, hard drives have the greatest potential to contain larger amounts of information. The reason for this statement is that with such large capacities available for a historically low price (approximately 0.015 cents / GB), there are many of large drives making their way on to the second hand market. The other issue is that to securely erase large drives takes a considerable amount of time (Valli & Patak, 2005).

This year saw the Australian study focused heavily on enterprise level small computer systems interface (SCSI) hard drives, with 2.5" notebook hard drives also examined. This bias was done with the intent of building a profile of large corporate and government sources, and this was again achieved as an outcome. The sources for the hard disk drives were a mixture of national on-line auctions and also traditional face to face auctions in the metropolitan area of Perth, Western Australia. This year 48 cases were examined. It should be noted that several of these were large RAID configurations.

The research on the disks was again undertaken using only tools that it was considered would be available to anyone who had obtained such disks. The series of actions were again an escalation of expertise in digital forensics. Should the hard disk not boot in suitable hardware, tools were used which carried out the same functions as the Windows Unformat and Undelete commands. In these cases a hex editor was also used to view any information that existed in the unallocated portion of the disk. For the extended analysis a customised Linux Debian install was used. In particular the scalpel file carver (Foremost 2007) and Autopsy the forensic browser (Sleuthkit 2006) were used in the analysis phase of disks that had been formatted.

As with previous years (Valli, 2004; Valli & Jones, 2005; Jones *et al*, 2006; Valli & Woodward, 2007), the first objective of the research was to determine whether there was any information on the disk that was readily recoverable with the techniques and tools identified above. The second stage of the research was to look for specific elements of information that would allow for the identification of the organisation or individual that had used the particular disk. Further, and, if possible, information such as the usernames, email addresses or documents, spreadsheets and data types of interest were examined. The purpose of this phase of the research was to determine the proportion of the disks that could be traced to an organisation or an individual and the level of exposure that data recovered would represent. The level of exposure looked not only at the privacy vector but also exposure that would allow the ready commissioning of criminal acts against the previous hard disk owner.

## OVERALL TRENDS UNCOVERED IN THE RESEARCH

Overall in the study once again a lack of pornographic material was uncovered given the hyperbole and claims made by some parties with respect to the use of the Internet to access this material during work hours is cast further into doubt. Unlike the 2006 study, there were no detected cases of child pornography in any of the analysed hard disks from Australia.

In 2008, a high percentage of the examined hard disks that yielded data, again contained significant personal and corporate exposure of confidential and commercial information. This phenomenon is a disturbing consistent trend in the studies so far.

This year, however, did see a slight increase in the amount of competently erased hard disks. This occurrence combined with the same observation in last years study that many of the hard disks were acquired from organisations with the ability and resources to erase the hard disks is gaining veracity. Potentially also with many of the disks coming from servers, there may be a recognised need to erase these before disposal. Drives purchased from what seemed to be low trade individuals were more likely to have not formatted the drive, or to have formatted but not erased.

2008004-AU 1.8 Terabyte RAID from a chemical supply company. There was the entire catalogue and ordering systems for the company. It also included a complete website for the company. The dates on the drive were less than 3 weeks old. There was also business documents, spreadsheets and other data you would expect to find in a medium sized enterprise.

2008013-AU This case contained 2 x 36GB SCSI drives that were extracted from a Sun Server. The server was sold as is and was advertised as being unable to boot. The drives were extracted and imaged. In the meantime it was possible to resurrect the server and boot it with other hard disk media. The original disks although imaged had not been examined they were placed back in the server, the server booted to a Solaris login prompt. The subsequent forensic analysis revealed that the drives were from a superannuation board as per the asset stickers that were on the external case of the server. This server yielded administrator passwords, configuration files and also network details.

2008015-AU This case was a 4 x 36GB RAID. There was an identifiable World of Warcraft avatar on the machine. There was a large number of pornographic thumbnails but they were all 100 x 100 GIF files, no other substantive files of this type were found. The thumbnails were predominately from mature age porn sites. There were no documents able to be located on the drive.

2008020-023-AU These four 72GB 68 pin SCSI drives were all unformatted and appeared to have originally belonged to a mining company. Analysis of the drives indicated that they were separate SCADA masters. As such they contained information including, the Historian database, iFix and other SCADA components relevant to the companies process control. In addition, passwords and usernames, including administrator, were still present on the drives. There were also photos of the site and personnel present.

2008026-028-AU These three 72GB SCSI drives had belonged to an organisation that provided IT support for Australian and international banks, as well as network services for other critical infrastructure providers. The drives had been formatted, but not erased and forensic recovery yielded a large range of information. Remote access policies and procedures for staff to access the banks networks, and network diagrams were amongst the information uncovered. Names of staff and what type of equipment they had been issued were also present.

2008030-AU This 146 GB SCSI drive had been formatted, but the data was recoverable. The drive was a Windows Server, and appeared to belong to a crop protection company. The server appeared to have been used for storing voicemail, and may also have been used for VoIP. Registry hives and other documentation relating to the companies business were also recovered.

2008036-AU This 72GB SCSI drive had been formatted but data was recoverable. The drive appeared to be a Windows server domain controller from a community health centre. It contained minutes from meetings and some other documentation. It also contained illegal software keys.

2008037-AU This 72GB SCSI drive had been formatted but data was recoverable. The drive belonged to a Nursing home, and contained extensive information. Patient information, letters from medical doctors, drug information, pictures of patients and their wounds, menus, accounts, minutes of meetings were a sample of the information recoverable

## POSSIBLE CAUSES

There is little chance that IT professionals could be unaware of the dangers of remnant data given the extensive coverage that this topic has now received. The availability of free and competent erasure tools such as DBAN also removes any financial impediment to acquisition of appropriate tools to accomplish erasure of media.

It is of considerable concern that so many enterprise level hard disks still contained recoverable data. Of greater concern is that some of this data was essential to the companies operations, and if made public could lead to significant civil and even criminal ramifications for the company. This year uncovered a set of drives that had a complete SCADA setup for a mine processing plant. This sort of exposure has several serious implications for security beyond the simple disclosure of the information or potential for theft. The intelligence found on the drive could make it a relatively trivial task to attack the plant that could result in a release of toxic chemicals or permanent destruction of the environment. From an economic perspective, destruction or partial destruction of the asset via deliberate malfeasance could be significant and widespread.

In the 2007 study, Valli and Woodward, postulated that the drives may have been purloined from organisations and on sold through the auction systems. The authors still contend that this maybe an option still being exercised by disgruntled employees this however does not explain the significant exposures uncovered. It may also be the case that organisations are relying, either implicitly or explicitly, on disposal companies / individuals to erase the drives for them. If this is the case, then it may be a failure in the processes of the disposer that has lead to this problem being so significant.

This year again it is clear that public and private organisations across a range of industry sectors are failing to discharge their responsibility to protect customer's details and sensitive data adequately. There is still no official mandated law or statute that requires organisations in Australia to erase secondary memory devices such as hard disk drives and this problem is growing. Evidence of inadequate protection uncovered in this and previous studies surely should now see this being considered by governments to protect citizens from themselves. Failing a legislated approach to the problem the creation for government based organisations of a centralised clearance service for the destruction/safe erasure of secondary storage media is needed.

It could also be safely deduced that organisations are failing to conduct adequate risk analysis of the issue of remnant data on secondary memory devices. It could not be similarly reasonably argued that the problem of remnant data is not a new or unknown phenomenon (Anonymous, 2003; de Paula, 2004; Duvall, 2003; Garfinkel & Shelat, 2003; Jones, 2006; Rohan, 2002; Spring, 2003; Valli, 2004; Valli & Jones, 2005; Valli & Patak, 2005). There are now a wide range of devices that are standard artifices of modern business these include, organisational servers, desktops, laptops, mobile phones, USB memory sticks, USB hard drives, even iPOD and MP3 player devices.

The risk versus return equation is simply not making sense for any modern organisation or individual who storage technologies. Auctioneers and sellers of these hard disks are also unwittingly providing potential criminals with targeted options for purchase. Many of the sellers have advertising that clearly indicates that the devices are from financial institutions, superannuation boards, insurance companies, mining companies to name a few. One has to ask what the provenance of these devices has to do with their suitability or value for use as storage mechanisms. There is some argument that these types of organisations drives may have been looked after in clean, regulated environments but that is where it would logically end.

## CONCLUSION

Similar to last years study, this year has again uncovered significant exposure of private, sensitive or fungible data on inadequately erased secondary memory devices. Organisations spend millions on protecting IT assets annually with firewalls, virus protection, intrusion prevention systems and other security countermeasures. They conduct regular audits of assets and procedures in an effort to protect and secure the information contained on their systems. We argue that these expenditures are largely symbolic and almost farcical when hard disks are disposed of without adequate protections.

The recovery of hard drives that contained complete process control system data and information is of considerable concern, as there are potentially lives at risk if the process control system is interfered with. The Australian Federal Government has the computer and network vulnerability assessment (CNVA) program and other initiatives and education programs in place to prevent such an event happening, which leads the authors to ponder as to whether the company in question had availed themselves of this program. The small office, home office (SOHO) users can be excused for disposing of drives that had been simply formatted, and not erased, but there is absolutely no excuse for multinationals to be disposing of unformatted hard drives with critical infrastructure information contained within. Possible legislation may include requiring critical infrastructure providers to undergo an external annual security audit to make sure that policies and standards are being adhered to.

This years study has revealed that not only is sensitive information being sold on the cheap, but so potentially are lives or the environment. The most expensive hard disk purchased in this years study was a 73GB SCSI hard drive that cost $85, with most costing less than that. The drives belonging to a mining company were purchased for just over $60, making them less than $1 per gigabyte. I don't see companies asking "Would you like private medical data, process control systems information, or network diagrams with that?", yet this is exactly what is still occurring today somewhere at an IT disposal sale or online auction in Australia.

In closing this study has been going in one form or another since 2004 it is now 2008 and still we are finding drives with significant exposures and realisable threats to individuals and the community at large. Some onus has to be placed on operating systems providers to enable a disk erasure routine for safe disposal of equipment. This study, has added further weight to the body of literature on the problem of remnant data. There is a real need for governments and in particular Australian governments to legislate and require organisations and custodians of electronic data to expunge it before disposal of the physical asset.

## REFERENCES

Foremost (2007) Foremost, retrieved 25$^{th}$ October 2008 from http://foremost.sourceforge.net/

Jones, A., Valli, C., Sutherland, I. and Thomas, P (2006). The 2006 Analysis of Information Remaining on Disks Offered for Sale on the Second Hand Market. *Journal of Digital Forensics, Security and Law, 1*(3), 23-36.

Sleuthkit (2006). Autopsy overview, retrieved 25$^{th}$ October 2008 from http://www.sleuthkit.org/autopsy/

Valli, C. (2004). *Throwing the Enterprise out with the Hard Disk.* In Proceedings of the 2nd Australian Computer, Information and Network Forensics Conference, Fremantle, Western Australia.

Valli, C., & Jones, A. (2005). *A UK and Australian Study of Hard Disk Disposal.* In Proceedings of the 3rd Australian Computer, Information and Network Forensics Conference, Edith Cowan University, Perth, Western Australia.

Valli, C., & Patak, P. (2005). *An investigation into the efficiency of forensic erasure tools for hard disk mechanisms.* In Proceedings of the 3rd Australian Computer, Information and Network Forensics Conference, Edith Cowan University, Perth, Western Australia.

Valli, C. & Woodward, A. (2007). Oops they did it again: The 2007 Australian study of remnant data contained on 2nd hand hard disks. In Proceedings of the 5th Australian Digital Forensics Conference, Edith Cowan University, Perth, Western Australia.

## COPYRIGHT